\pdfoutput=1
\documentclass[aps,12pt]{revtex4-1}

\usepackage{mathtools}
\usepackage{amssymb}
\usepackage{amsmath}
\usepackage{hyperref}
\usepackage{feynmp-auto}

\newcommand\dlr{\raisebox{0.1em}{$\stackrel{\scriptstyle\leftrightarrow}\partial$}}
\renewcommand\d{\partial}

\newcommand\+\dagger

\begin{document}

\title{Effective field theory for one-dimensional nonrelativistic particles with contact interaction}
\author{Soke Yuen Yong and Dam Thanh Son}
\affiliation{Kadanoff Center for Theoretical Physics and Department of Physics, University of Chicago, Illinois 60637, USA}

\begin{abstract}
  We consider a field theory describing interacting nonrelativistic
  particles of two types, which map to each other under time reversal,
  with point-like interaction.  We identify a new type of interaction
  which depends on the relative velocity between the particles.  We
  compute the renormalization group running of the coupling constants
  and find a fixed point and a fixed line. We show that the scattering
  amplitudes can be expressed in terms of three parameters.  The
  result matches with a quantum mechanical analysis and represents the
  most general point-like interaction consistent with unitarity and
  time reversal invariance.
\end{abstract}

\maketitle

\section{Introduction}

Effective field theory has become a useful tool to investigate
nonrelativistic systems, from nucleons to trapped
atoms~\cite{Weinberg:1990rz,Weinberg:1991um,Kaplan:1998tg,Braaten:2004rn}.
Effective field theory provides a framework to parametrize low-energy
properties of such systems in terms of a small number of free
parameters appearing in the Lagrangian.  The technique is very useful
in the case of fermions at unitarity \cite{OHara:2002pqs,Regal:2004zza,Zwierlein:2004zz}, where field-theory techniques
have given additional insights to quantum mechanical problems.  An
example is the analysis of the Efimov effect using renormalization
group~\cite{Bedaque:1998kg,Bedaque:1998km,Braaten:2004rn}.

In this paper, we apply effective field theory to the problem of
nonrelativistic particles in one spatial dimension.  The reason for us
to revisit this seemingly trivial problem is the recent
suggestion~\cite{Geracie:2016dti,Geracie:unpublished} of a new
Galilean-invariant interaction between nonrelativistic particles.  In
one spatial dimension, such an interaction is a pair-wise local interaction
with strength proportional to the relative velocity of the
participating particles.

We are also motivated by the possibility of finding new
one-dimensional systems with nonrelativistic conformal
symmetry~\cite{Nishida:2007pj}, similar to fermions at unitarity in
three dimensions and anyons in two dimensions.  In one spatial
dimension, the only theory found so far with this symmetry is the
theory with four-particle contact interaction~\cite{Nishida:2009pg}.
In field theory, conformal field theory emerges at RG fixed points,
thus RG is the most convenient method to find such theories.  Another
motivation is to give an effective field theory interpretation of an
extensive mathematical literature on self-adjoint extension of the 1D
Hamiltonian (see Refs.~\cite{Coutinho:1997,Albeverio:2004} and
references therein).

The structure of the paper is as follows.  In Sec.~\ref{sec:L}, we
consider a theory of two distinct species of particles which map
to each other under parity. We construct a Lagrangian
consistent with Galilean invariance containing only up to two
derivatives.  In Sec.~\ref{sec:A} we evaluate the full scattering
amplitudes (the transmission and reflection amplitudes).  Then in
Sec.~\ref{sec:RG} we develop a renormalization group treatment of the
two-body sector of the theory and find that the theory has a fixed
point and a fixed line.  We find the general form of scattering
amplitudes, and make contact with quantum-mechanical calculations.  In
Sec.~\ref{sec:cases} we describe the physics at the fixed points and
in some special cases of the RG flow.  Section \ref{sec:concl}
contains concluding remarks.  The Appendix contains a purely quantum
mechanical treatment of the problem.

\section{Lagrangian}
\label{sec:L}

We consider two species of nonrelativistic particles, $\psi_1$ and
$\psi_2$, living in one spatial dimension and interacting though a
point-like interaction. The Lagrangian that we consider is
\begin{multline}\label{eq:L}
\mathcal L =
   \sum_{a=1}^2 \left(
   i \psi^\dagger_a\d_t\psi_a
    - \frac{1}{2m_a} \partial_x\psi^\dagger_a \partial_x\psi_a \right)
	- \lambda \psi^\dagger_1\psi_1 \psi^\dagger_2\psi_2\\
	+\frac i2 a
	(m_1 \psi^\dagger_1\psi_1 \psi^\dagger_2\dlr_x\psi_2
	-m_2 \psi^\dagger_2\psi_2 \psi^\dagger_1\dlr_x\psi_1) 
	- \frac c4\, (\psi^\dagger_2\dlr_x\psi^\dagger_1) (\psi_2\dlr_x\psi_1).
\end{multline}

The theory does not generally have a parity symmetry $x\to-x$.
However when $m_1=m_2$, it is invariant under the combination of
$x\to-x$ and $\psi_1\leftrightarrow\psi_2$.  The same is true for the
combination of time reversal and $\psi_1\leftrightarrow\psi_2$.
It is easy to see also that the theory has Galilean symmetry.  
The most nontrivial check is for the interaction term proportional to
$a$, which, in the first-quantized language, can be written as $m_1m_2
a\{v-v',\,\delta(x-x')\}$, where $v$ and $v'$ are the velocities of the interacting particles.  This interaction is obviously invariant under Galilean
boosts.

From naive power counting, $\lambda$ has dimension 1 (i.e., that of
momentum), $a$ is dimensionless, and $c$ has dimension $-1$.  One may
ask why one includes in the Lagrangian an irrelevant interaction.  The
reason, as we see in the next section, is that there exists a
nonperturbative fixed point for $c$, in exactly the same manner as the
naively irrelevant four-point interaction in 3D nonrelativistic theory
has a UV fixed point.  As we also see in the next section, the
Feynman diagrams in theory (\ref{eq:L}) contains only one linear
divergence, and this special property is not preserved if other
interaction terms are added int the Lagrangian.

\section{Scattering amplitudes}
\label{sec:A}

\unitlength=1.2mm
\begin{fmffile}{four-point-vertex}
\begin{figure}
        \centering
        \begin{fmfgraph*}(40,25)
                \fmfleft{i1,i2}
                \fmfright{o1,o2}
                \fmflabel{$\psi_2$}{i1}
                \fmflabel{$\psi_1$}{i2}
                \fmflabel{$\psi^\dagger_2$}{o1}
                \fmflabel{$\psi^\dagger_1$}{o2}
                \fmf{fermion,label=$\overrightarrow{p_3}$}{i1,v1}
                \fmf{fermion,label=$\overrightarrow{p_2}$}{v1,o2}
                \fmf{fermion,label=$\overrightarrow{p_1}$}{i2,v2}
                \fmf{fermion,label=$\overrightarrow{p_4}$}{v2,o1}
                \fmf{photon}{v1,v2}
        \end{fmfgraph*}
        \vspace*{7mm}
        \caption{Four-particle interaction vertex}
        \label{fig:interactionvertex}
\end{figure}
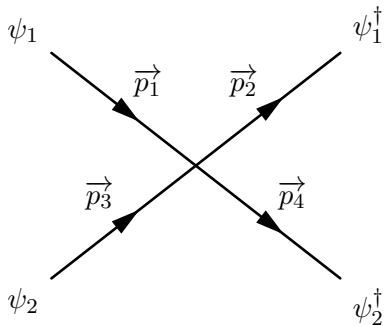
\end{fmffile}
To compute the scattering amplitude, we first write down the vertex
presented in Fig.~\ref{fig:interactionvertex},
\begin{equation}
  V(p_1,p_3;p_2,p_4) = -i\lambda - \frac i2 a[m_1(p_3+p_4)-m_2(p_1+p_2)]
  -\frac i4 c (p_1-p_3)(p_2-p_4).
\end{equation}
In further calculations we will set $m_1=m_2=1$.

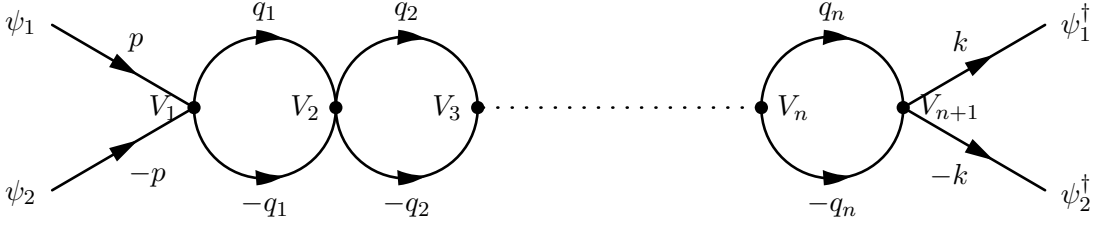
\begin{figure}
                \centering
                \begin{fmfgraph*}(150,20)
                        \fmfleft{i1,i2}
                        \fmfright{o1,o2}
                        \fmf{fermion,label=$-p$}{i1,v1}
                        \fmf{fermion,label=$p$}{i2,v1}
                        \fmf{fermion,label=\(q_1\),left}{v1,v2}
                        \fmf{fermion,label=\(-q_1\),right}{v1,v2}
                        \fmf{fermion,label=\(q_2\),left}{v2,v3}
                        \fmf{fermion,label=\(-q_2\),right}{v2,v3}
                        \fmf{dots}{v3,v4}
                        \fmf{fermion,label=\(q_n\),left}{v4,v5}
                        \fmf{fermion,label=\(-q_n\),right}{v4,v5}
                        \fmf{fermion,label=$-k$}{v5,o1}
                        \fmf{fermion,label=$k$}{v5,o2}
                        \fmfdot{v1,v2,v3,v4,v5}
                        \fmflabel{$\psi_2$}{i1}
                        \fmflabel{$\psi_1$}{i2}
                        \fmflabel{$\psi^\dagger_2$}{o1}
                        \fmflabel{$\psi^\dagger_1$}{o2}
                        \fmfv{label=\(V_1\)}{v1}
                        \fmfv{label=\(V_2\)}{v2}
                        \fmfv{label=\(V_3\)}{v3}
                        \fmfv{label=\(V_n\)}{v4}
                        \fmfv{label=\(V_{n+1}\)}{v5}
                \end{fmfgraph*}
                \vspace*{1mm}
                \caption{Diagram with \(n\) loops and \(n+1\) vertices. }
                \label{fig:nloop}
        \end{figure}

The total scattering amplitude is given by the sum of diagrams such as
in Figure \ref{fig:nloop}, for all possible numbers of loops.
In the center-of-momentum frame, the incoming particles have
momenta $\pm p$, and the total energy is $E$. If the particles
are on-shell, then $E=p^2$.  A diagram with $n$ loops (and therefore
$n+1$ vertices, see Figure \ref{fig:nloop}), contributes to the total
amplitude a term of the form
\begin{equation}
A_{n+1} =
\!\int\! \prod_{i=1}^n  \frac{d\omega_i\, dq_i}{(2 \pi)^2}\,
 \prod_{i=1}^n\frac{i}{(\frac{E}{2} + \omega_i - \frac{q_i^2}{2} + i \epsilon)}\,
 \frac{i}{(\frac{E}{2} - \omega_i - \frac{q_i^2}{2} + i \epsilon)}
 \prod_{i=1}^{n+1} V_i V_{n+1},
\end{equation}
where $V_i =V(q_{i-1},q_i)$ is the vertex factor of the $i$th vertex, and where
$q_0\equiv p$ and $q_{n+1}=k$.

\unitlength=1.1mm
\begin{fmffile}{n-bubble}
\end{fmffile}

It is clear from Fig.~\ref{fig:nloop} and the structure of the
interaction vertex that in the product of $V$'s, each $q_i$ can only
appear to the power of 0, 1, or 2.  Thus, we only need to compute the
integral
\begin{equation}
  B_n = \!\int\! \frac{d\omega\, dq}{(2 \pi)^2}\,
  \frac{i}{\frac{E}{2} + \omega - \frac{q^2}{2} + i \epsilon}\,
  \frac{i}{\frac{E}{2} - \omega - \frac{q^2}{2} + i \epsilon}
  q^n 
  = \int\! \frac{dq}{2 \pi}\,
  \frac{iq^n}{E-q^2+i\epsilon} \,.
\end{equation}
Obviously $B_1=0$.  The integral for $B_0$ converges 
\begin{equation}\label{eq:B0}
  B_0 =\frac1{2\sqrt{E}}\,,
\end{equation}
and that for $B_2$ diverges linearly.  Using a sharp momentum cutoff
$|q|<\Lambda$, the latter is 
\begin{equation}\label{eq:B2}
B_2 =  -\frac{i\Lambda}\pi + \frac{\sqrt{E}}2 \,.
\end{equation}

Therefore, only combinations of vertex factors
$V_1V_2...V_{n+1}$ that have $q_i$ to the power of 0 or 2 for all
$i$ will produce nonzero contributions to the total scattering
amplitude. That is, if one particular vertex involves a factor of
$q_i$, then the next vertex has to have the same factor, and the
combination of the two vertices results in a factor of $B_2$ in the
scattering amplitude term corresponding to that particular diagram.

The full scattering amplitude satisfies the recursion relation
\begin{equation}\label{eq:AV-recursion}
  A(E; p,k) = V(p,k) + \int\!\frac{d\omega\,dq}{(2\pi)^2}\,
  V(p,q) \frac{i}{\frac{E}{2} + \omega - \frac{q^2}{2} + i \epsilon}
  \frac{i}{\frac{E}{2} - \omega - \frac{q^2}{2} + i \epsilon}
  A(E; p,k),
\end{equation}
where
\begin{equation}
  V(p,k)\equiv V(p,-p;k,-k) = -i \lambda + i a (p+k) - i cpk .
\end{equation} 
To solve this equation, we note
that the external momenta $p$ and $k$ can appear, through the
left-most and right-most vertices, in powers 0 or 1 in the scattering
amplitude.  We thus can make the following ansatz
\begin{equation}\label{eq:ansatz}
  A (E;p,k) = A^{0}(E) + A^{1}(E) p + A^{2}(E) k + A^{3}(E) pk .
\end{equation}
Here the $A^{i}$, $i=0,1,2,3$ are coefficients that depend on $E$, but
not on $p$ and $k$.  Substituting the ansatz (\ref{eq:ansatz}) into
the recursion relation (\ref{eq:AV-recursion}), one finds
\begin{equation}
\begin{pmatrix}
A^{0} \\ A^{1} \\ A^{2} \\ A^{3}
\end{pmatrix}
= \begin{pmatrix} -i\lambda\\ ia \\ ia \\ -ic \end{pmatrix} +
\begin{pmatrix}
-i \lambda B_0 & 0 & i a B_2 & 0 \\
0 & -i \lambda B_0 & 0 & i a B_2 \\
i a B_0 & 0 & -i c B_2 & 0 \\
0 & i a B_0 & 0 & - i c B_2 \\
\end{pmatrix}
\begin{pmatrix}
A^{0} \\ A^{1} \\ A^{2} \\ A^{3}
\end{pmatrix} .
\end{equation}
The solution to this linear system of equations is
\begin{equation}
\begin{pmatrix}
  A^{0} \\  A^{1} \\ A^{2} \\ A^{3}
\end{pmatrix}
= \frac{1}{(1 + i\lambda B_0)(1+icB_2) +  a^2 B_0B_2} 
\begin{pmatrix}
-i \lambda + (c\lambda - a^2) B_2 \\ 
i  a \\ 
i  a \\ 
- i c + (c\lambda -a^2) B_0
\end{pmatrix}.
\end{equation}
Therefore,
\begin{align}
  A(p,k) &= \frac{-i \lambda  + (c\lambda-a^2)(B_2+B_0pk) + ia(p+k)
    - i c  pk}
{(1 + i\lambda B_0)(1+icB_2) +  a^2 B_0B_2} \,.
\end{align}

For forward scattering, $k = p$, and so the forward scattering amplitude is
\begin{equation}\label{eq:Af}
   A_F=\frac{-i\lambda + (c\lambda -a^2) (B_2 + B_0p^2)
	+2 i ap - i cp^2}
{(1 + i\lambda B_0)(1+icB_2) +  a^2 B_0B_2} \,.
\end{equation}
Backward scattering occurs when $k = -p$, and so the backward
scattering amplitude is
\begin{equation}\label{eq:Ab}
  A_B=\frac{-i\lambda + (c\lambda -a^2) (B_2 - B_0p^2)
	 + i cp^2}
  {(1 + i\lambda B_0)(1+icB_2) +  a^2 B_0B_2} \,.
\end{equation}	
In these equation $B_0$ and $B_2$ should be evaluated according to
Eqs.~(\ref{eq:B0}) and (\ref{eq:B2}) at the onshell value of the
incoming energy $E=p^2$.
	
\section{Renormalization group analysis}
\label{sec:RG}

The amplitudes computed in the previous section depends on the
coupling constant $\lambda$, $a$, $c$, and the momentum cutoff
$\Lambda$.  We now write down the renormalization group equation for
the coupling constants, demanding that  physical quantities (in
this case, the scattering amplitudes) are invariant when one
simultaneously changes the cutoff and the couplings.  Introducing the
dimensionless coupling constants $\tilde\lambda =
\Lambda^{-1}\lambda$, $\tilde a=a$, $\tilde c=\Lambda c$, the
renormalization group equations can be obtained from the condition
\begin{equation}\label{eq:GL}
  \left(\Lambda \frac\d{\d \Lambda}
    + \beta_{\tilde\lambda}\frac\d{\d\tilde\lambda}
    + \beta_{\tilde a} \frac\d{\d\tilde a}
    + \beta_{\tilde c} \frac\d{\d\tilde c}\right)
  A_{F,B}(\Lambda, \tilde\lambda, \tilde a, \tilde c) = 0.
\end{equation}
From this equation, which should be satisfies for all $p$, one obtains
the beta-functions
\begin{subequations}
\begin{align}
   \beta_{\tilde\lambda} &= -\tilde\lambda + \frac{\tilde a^2}\pi\, ,\\
   \beta_{\tilde a} &= \frac{\tilde a \tilde c}\pi\,,\\
   \beta_{\tilde c} &= \frac{\tilde c^2}\pi + \tilde c.
\end{align}
\end{subequations}
Since the scattering amplitudes are exact, the beta functions are also
exact.  The fact that Eq.~(\ref{eq:GL}) can be satisfied for all $p$
means that no additional interaction is generated by the RG apart from the
ones that have been included in the Lagrangian.

Setting the beta functions to zero, we find that our theory has an isolated
nontrivial fixed
point at $(\tilde\lambda, \tilde a, \tilde c) = (0,0,-\pi)$,
as well as a line of fixed points at $(\tilde\lambda,
\tilde a, \tilde c) = (\frac{1}\pi \tilde{a}^2, \tilde a,0)$.  The
isolated fixed points have three relevant directions, while points on
the fixed line have one relevant, one irrelevant and one marginal
directions.

The most general solution to the RG equation $\Lambda(\d
g_i/\d\Lambda)=\beta_{g_i}$ with $g_i=\tilde c,\tilde a,\tilde\lambda$
can be written as
\begin{subequations}
\begin{align}
  \tilde c(\Lambda) &= - \frac {2\pi\Lambda}{2\Lambda+\pi\mu} \,,\\
  \tilde a(\Lambda) &= \frac{2\pi\alpha\mu}{2\Lambda+\pi\mu} \,,\\
  \tilde \lambda(\Lambda) &=  \frac{4\alpha^2\mu}{2\Lambda+\pi\mu}
  + \frac{2\gamma}\Lambda \,.
\end{align}
\end{subequations}
where $\mu$, $\alpha$, and $\gamma$ are integration constants.
Note that $\mu$ and $\gamma$ have dimension of momentum while
$\alpha$ is dimensionless.

When $\gamma=0$ the above flow interpolate between two fixed points:
the isolated fixed point in the UV and a point on the fixed line in
the IR.  If $\gamma\neq0$, in the IR $\lambda$ blows up to $\pm\infty$
depending on the sign of $\gamma$.  An interesting regime is
$\mu\gg\gamma$.  Then one first flows from the isolated fixed point
($\Lambda\gg\mu$) to a point on the fixed line with $\tilde
a=2\alpha$, lingers there for $\Lambda$ in the interval
$\mu\gg\Lambda\gg\gamma$, and then $\tilde\lambda\to\pm\infty$ when
$\Lambda$ drops below $\gamma$.

The invariance of the scattering amplitudes with respect to variation
of the cutoff scale $\Lambda$ implies that these amplitudes can be
expressed completely in terms of the integration constants $\mu$,
$\alpha$, and $\gamma$.  In other words, the short-range interaction
can be completely characterized by the parameters $\mu$, $\alpha$, and
$\gamma$.  In terms of these parameters, the forward and backward
scattering amplitudes are
\begin{subequations}
\begin{align}
  A_F &= -2p\, \frac{p^2 +2 (\alpha\mu+i\alpha^2\mu +i\gamma)p -\gamma\mu}
  {p^2 + i ( \mu+\alpha^2\mu + \gamma) p - \gamma\mu} \,,\\
  A_B &= 2p\, \frac{ p^2 + \gamma\mu}
  {p^2 + i (\mu  +\alpha^2\mu + \gamma)p -\gamma\mu} \,,
\end{align}
\end{subequations}
which corresponds to the transmission and reflection coefficients:
\begin{subequations}\label{general-TR}
\begin{align}
  T &= 1+ \frac{A_F}{2p}= \frac{i[(1+i\alpha)^2\mu-\gamma]p}
  {p^2 + i ( \mu+\alpha^2\mu + \gamma) p - \gamma\mu} \,,\\
  R &= \frac{A_B}{2p} =\frac{ p^2 + \gamma\mu}
  {p^2 + i (\mu  +\alpha^2\mu + \gamma)p -\gamma\mu}  \,.
\end{align}
\end{subequations}
This coincides with the form of the scattering amplitudes derived in the Appendix, Eqs.~(\ref{eq:TR-app}), by solving a
quantum-mechanical problem of scattering from a point-like potential invariant under the combination
of time reversal and particle exchange.  As explained in the Appendix, such a potential is characterized by three complex numbers possessing the same phase.  In our case, the three complex numbers are
\begin{align}
  \mathcal A & = \frac{(1+\alpha^2)\mu+\gamma}{(1+i\alpha)^2\mu-\gamma} \,,\\
  \mathcal B & = \frac2{(1+i\alpha)^2\mu-\gamma} \,,\\
  \mathcal C & = \frac{2\gamma\mu}{(1+i\alpha)^2\mu-\gamma} \,.
\end{align}
They satisfy the condition $|\mathcal A|^2-|\mathcal B||\mathcal
C|=1$, required by charge conservation and time reversal.

\section{Special cases}
\label{sec:cases}

We now investigate the behavior of the scattering amplitudes in special cases.

\subsection{Generic infrared behavior}

Without any fine tuning, the behavior of the scattering amplitudes in
the IR can be obtained by setting $p\to0$ in Eqs.~(\ref{general-TR}):
\begin{equation}
  T = 0, \quad R = -1,
\end{equation}
corresponding to total reflection.  The minus sign in $R$ can be
understood by putting it into correspondence to the wave function of
the 1D scattering problem
\begin{equation}
  \psi(x) \sim \theta(-x) \sin kx .
\end{equation}
where $\theta(x)$ is the step function.
Such a wave function corresponds to scattering off a hard wall.  It is
also the behavior of the scattering amplitude in the IR for scattering
on a delta-function potential (see below)

\subsection{Delta-function potential}

The familiar problem of scattering on a delta-function potential
corresponds to $c=a=0$.  In this case the transmission and reflection
amplitudes can be obtained from Eqs.~(\ref{eq:Af}) and (\ref{eq:Ab}):
\begin{equation}
  T = \frac{2p}{2p+i\lambda} \,, \qquad R = -\frac{i\lambda}{2p+i\lambda}\,.
\end{equation}
In the infrared $p\to0$ we reproduce reflection on a hard wall, $T=0$
and $R=-1$.

\subsection{Fixed points}

Now we consider the behavior of the scattering amplitudes when the
coupling constants are at one of the fixed points.  Scale invariance
dictates, then, that the scattering amplitudes are
momentum-independent.

Consider the isolated fixed point.  We can approach the fixed
point by setting $\gamma=0$ and then taking the limit $p\gg\mu$.  Then
\begin{equation}
  R=1, \qquad T=0.
\end{equation}
At the fixed point we have total reflection but, in contrast to the
case of a hard-wall potential, the phase reflection amplitude is $1$
instead of $-1$. 

Now consider the system at a fixed point along the fixed line.  Again
we set $\gamma=0$ but now $p\ll\mu$.  One finds
\begin{equation}
  T = \frac{1+ i\alpha}{1-i\alpha}\,, \qquad R =0, 
\end{equation}
which implies total transmission.  The transmission amplitude, however, has a
nonzero phase shift, $T=e^{i\delta}$, with
\begin{equation}
  \delta = 2 \arctan\alpha .
\end{equation}

\subsection{Special RG flows}

One can investigate the whole flow from the isolated fixed point to a
point on the fixed line.  Such a flow requires fine-tuning $\gamma$ to
0.  One finds then
\begin{equation}
  T = \frac{i(1+i\alpha)^2\mu}{p+i(1+\alpha^2)\mu}
    = \frac{i(1+\alpha^2)\mu}{p+i(1+\alpha^2)\mu} e^{i\delta} \,, \qquad
  R = \frac p{p+i(1+\alpha^2)\mu}\,.
\end{equation}
At $\alpha=0$, this is the scattering off a ``$\delta'(x)$
potential,'' defined as in Ref.~\cite{Coutinho:1997}.  For
$\alpha\neq0$, the transmission amplitude is modified by a constant
phase.

We can investigate the RG flow along $c=0$.  For $\gamma\ne0$, this is
a flow between a point on the fixed line and $\gamma=\pm\infty$.  For
that we set $\mu\gg\gamma,\,p$.  The amplitudes become
\begin{equation}
  T = \frac{i(1+i\alpha)^2 p}{i(1+\alpha^2)p-\gamma}\,,\qquad
  R = \frac{\gamma}{i(1+\alpha^2)p -\gamma} \,,
\end{equation}
which can be rewritten as
\begin{equation}
  T = \frac{2p}{2p+i\tilde\lambda} e^{i\delta}, \qquad
  R = -\frac{i\tilde\lambda}{2p+i\tilde\lambda}\,,
\end{equation}
where
\begin{equation}
  \tilde\lambda = \frac{2\gamma}{1+\alpha^2} \,.
\end{equation}
The scattering amplitudes thus have the same form as those of a
delta-function potential, except that the transmission amplitude is
multiplied by a momentum-independent phase.

\section{Conclusion}
\label{sec:concl}

Using field-theory methods, we have analyzed a model involving
nonrelativistic particles of two types interacting through a contact
interaction.  A novel feature of the model is the presence of a
velocity-dependent interaction, which is still consistent with parity
and time reversal if these symmetries also exchange the two types of
particles.

Analyzing the RG group equations, we find that the theory has an
isolated fixed point and a line of fixed points.  At the fixed line,
the scattering amplitudes are trivial, up to a constant phase in the
transmission amplitude.  In the general case, we find that the
scattering amplitudes depend on three parameters, in perfect
correspondence with the quantum mechanical analysis.

It would be interesting to explore the few-body and many-body aspects
of the system.  We defer this problem to future work.

\acknowledgments

We thank Michael Geracie for sharing with us an unpublished note
\cite{Geracie:unpublished} which stimulated this work.  This work is
supported, in part, by ARO MURI grant No.\ 63834-PH-MUR, the Chicago
MRSEC, which is funded by NSF through Grant No. DMR-1420709, and a
Simons Investigator award by the Simons foundation.

\appendix

\section{Quantum mechanical analysis}
\label{sec:qm}

Denote the wavefunction of a system of two particles as $\psi(x,y)$,
where $x$ is the coordinate of the particle of species 1 and $y$ the
coordinate of the particle of species 2.  Denote by $z=x-y$ the
relative coordinate of the two particles.  The most general
interaction between particles at $x=y$ can be parametrized by a
boundary condition across $z=0$~\cite{Coutinho:1997}:
\begin{equation}\label{eq:bc}
  \begin{pmatrix} \psi\\ \psi' \end{pmatrix}_{z=+0} =
  \begin{pmatrix} \mathcal A & \mathcal B \\ \mathcal C & \mathcal D \end{pmatrix} \begin{pmatrix} \psi\\ \psi' \end{pmatrix}_{z=-0} ,
\end{equation}
where $\mathcal A$, $\mathcal B$, $\mathcal C$, and $\mathcal D$ are
complex number characterizing the short-range interaction.  Particle number conservation requires
that for two arbitrary wave functions $\psi_1$ and $\psi_2$, the matrix element of the particle number current between the two states,
$\psi_1^*\d_x\psi_2-\d_x\psi_1^*\psi_2$, is continuous across $z=0$.
This condition implies that
\begin{subequations}\label{eq:conditions}
\begin{align}
  \mathcal A^* \mathcal D - \mathcal C^* \mathcal B & = 1, \\
  \mathcal A^* \mathcal C - \mathcal C^* \mathcal A &= 0, \\
  \mathcal B^* \mathcal D - \mathcal D^* \mathcal B &= 0.
\end{align}
\end{subequations}
The last two equations imply that $\mathcal A^* \mathcal C$ and
$\mathcal B^* \mathcal D$ are real.

Time-reversal invariance, with exchanging the type of particles,
implies that $\tilde\psi(z)=\psi^*(-z)$ should also satisfy the boundary
condition (\ref{eq:bc}).  This implies that
\begin{equation}
  \begin{pmatrix} \mathcal A^* & -\mathcal B^* \\ -\mathcal C^* & \mathcal D^* \end{pmatrix} =
  \begin{pmatrix} \mathcal A & \mathcal B\\ \mathcal C & \mathcal D \end{pmatrix}^{-1} .
\end{equation}
From this condition and Eqs.~(\ref{eq:conditions}) one finds $\mathcal
A=\mathcal D$ and that $\mathcal A$, $\mathcal B$, and $\mathcal C$
have the same phase, i.e.,
\begin{equation}
  \begin{pmatrix} \mathcal A & \mathcal B \\ \mathcal C & \mathcal D \end{pmatrix} =
  \begin{pmatrix} A & B \\ C & A \end{pmatrix} e^{i\theta} ,
\end{equation}
where $A$, $B$, and $C$ are real numbers satisfying the equation
\begin{equation}
  A^2 - BC = 1.
\end{equation}
and $\theta$ is also real.
Thus, a short-range interaction is characterized by two real numbers
and one phase.

The transmission and reflection amplitudes can be computed by
requiring the wavefunction
\begin{equation}
  \psi(z) = \left\{ \begin{array}{lll} e^{ipz} + R e^{-ipz}, & & z<0, \\ T e^{ipz}, & & z>0, \end{array} \right. 
\end{equation}
to satisfy the boundary condition (\ref{eq:bc}), which yields the result
\begin{equation}\label{eq:TR-app}
  T = \frac{-2ip e^{i\delta}}{C-2iA p-Bp^2}\,, \qquad
  R = -\frac{C+Bp^2}{C-2iAp -Bp^2}\,.
\end{equation}

\bibliography{1Dmodel}	

\begin{thebibliography}{15}%
\makeatletter
\providecommand \@ifxundefined [1]{%
 \@ifx{#1\undefined}
}%
\providecommand \@ifnum [1]{%
 \ifnum #1\expandafter \@firstoftwo
 \else \expandafter \@secondoftwo
 \fi
}%
\providecommand \@ifx [1]{%
 \ifx #1\expandafter \@firstoftwo
 \else \expandafter \@secondoftwo
 \fi
}%
\providecommand \natexlab [1]{#1}%
\providecommand \enquote  [1]{``#1''}%
\providecommand \bibnamefont  [1]{#1}%
\providecommand \bibfnamefont [1]{#1}%
\providecommand \citenamefont [1]{#1}%
\providecommand \href@noop [0]{\@secondoftwo}%
\providecommand \href [0]{\begingroup \@sanitize@url \@href}%
\providecommand \@href[1]{\@@startlink{#1}\@@href}%
\providecommand \@@href[1]{\endgroup#1\@@endlink}%
\providecommand \@sanitize@url [0]{\catcode `\\12\catcode `\$12\catcode
  `\&12\catcode `\#12\catcode `\^12\catcode `\_12\catcode `\%12\relax}%
\providecommand \@@startlink[1]{}%
\providecommand \@@endlink[0]{}%
\providecommand \url  [0]{\begingroup\@sanitize@url \@url }%
\providecommand \@url [1]{\endgroup\@href {#1}{\urlprefix }}%
\providecommand \urlprefix  [0]{URL }%
\providecommand \Eprint [0]{\href }%
\providecommand \doibase [0]{http://dx.doi.org/}%
\providecommand \selectlanguage [0]{\@gobble}%
\providecommand \bibinfo  [0]{\@secondoftwo}%
\providecommand \bibfield  [0]{\@secondoftwo}%
\providecommand \translation [1]{[#1]}%
\providecommand \BibitemOpen [0]{}%
\providecommand \bibitemStop [0]{}%
\providecommand \bibitemNoStop [0]{.\EOS\space}%
\providecommand \EOS [0]{\spacefactor3000\relax}%
\providecommand \BibitemShut  [1]{\csname bibitem#1\endcsname}%
\let\auto@bib@innerbib\@empty
\bibitem [{\citenamefont {Weinberg}(1990)}]{Weinberg:1990rz}%
  \BibitemOpen
  \bibfield  {author} {\bibinfo {author} {\bibfnamefont {S.}~\bibnamefont
  {Weinberg}},\ }\href {\doibase 10.1016/0370-2693(90)90938-3} {\bibfield
  {journal} {\bibinfo  {journal} {Phys. Lett.}\ }\textbf {\bibinfo {volume}
  {B251}},\ \bibinfo {pages} {288} (\bibinfo {year} {1990})}\BibitemShut
  {NoStop}%
\bibitem [{\citenamefont {Weinberg}(1991)}]{Weinberg:1991um}%
  \BibitemOpen
  \bibfield  {author} {\bibinfo {author} {\bibfnamefont {S.}~\bibnamefont
  {Weinberg}},\ }\href {\doibase 10.1016/0550-3213(91)90231-L} {\bibfield
  {journal} {\bibinfo  {journal} {Nucl. Phys.}\ }\textbf {\bibinfo {volume}
  {B363}},\ \bibinfo {pages} {3} (\bibinfo {year} {1991})}\BibitemShut
  {NoStop}%
\bibitem [{\citenamefont {Kaplan}\ \emph {et~al.}(1998)\citenamefont {Kaplan},
  \citenamefont {Savage},\ and\ \citenamefont {Wise}}]{Kaplan:1998tg}%
  \BibitemOpen
  \bibfield  {author} {\bibinfo {author} {\bibfnamefont {D.~B.}\ \bibnamefont
  {Kaplan}}, \bibinfo {author} {\bibfnamefont {M.~J.}\ \bibnamefont {Savage}},
  \ and\ \bibinfo {author} {\bibfnamefont {M.~B.}\ \bibnamefont {Wise}},\
  }\href {\doibase 10.1016/S0370-2693(98)00210-X} {\bibfield  {journal}
  {\bibinfo  {journal} {Phys. Lett.}\ }\textbf {\bibinfo {volume} {B424}},\
  \bibinfo {pages} {390} (\bibinfo {year} {1998})},\ \Eprint
  {http://arxiv.org/abs/nucl-th/9801034} {arXiv:nucl-th/9801034} \BibitemShut
  {NoStop}%
\bibitem [{\citenamefont {Braaten}\ and\ \citenamefont
  {Hammer}(2006)}]{Braaten:2004rn}%
  \BibitemOpen
  \bibfield  {author} {\bibinfo {author} {\bibfnamefont {E.}~\bibnamefont
  {Braaten}}\ and\ \bibinfo {author} {\bibfnamefont {H.~W.}\ \bibnamefont
  {Hammer}},\ }\href {\doibase 10.1016/j.physrep.2006.03.001} {\bibfield
  {journal} {\bibinfo  {journal} {Phys. Rept.}\ }\textbf {\bibinfo {volume}
  {428}},\ \bibinfo {pages} {259} (\bibinfo {year} {2006})},\ \Eprint
  {http://arxiv.org/abs/cond-mat/0410417} {arXiv:cond-mat/0410417} \BibitemShut
  {NoStop}%
\bibitem [{\citenamefont {O'Hara}\ \emph {et~al.}(2002)\citenamefont {O'Hara},
  \citenamefont {Hemmer}, \citenamefont {Gehm}, \citenamefont {Granade},\ and\
  \citenamefont {Thomas}}]{OHara:2002pqs}%
  \BibitemOpen
  \bibfield  {author} {\bibinfo {author} {\bibfnamefont {K.~M.}\ \bibnamefont
  {O'Hara}}, \bibinfo {author} {\bibfnamefont {S.~L.}\ \bibnamefont {Hemmer}},
  \bibinfo {author} {\bibfnamefont {M.~E.}\ \bibnamefont {Gehm}}, \bibinfo
  {author} {\bibfnamefont {S.~R.}\ \bibnamefont {Granade}}, \ and\ \bibinfo
  {author} {\bibfnamefont {J.~E.}\ \bibnamefont {Thomas}},\ }\href {\doibase
  10.1126/science.1079107} {\bibfield  {journal} {\bibinfo  {journal}
  {Science}\ }\textbf {\bibinfo {volume} {298}},\ \bibinfo {pages} {2179}
  (\bibinfo {year} {2002})},\ \Eprint {http://arxiv.org/abs/cond-mat/0212463}
  {arXiv:cond-mat/0212463} \BibitemShut {NoStop}%
\bibitem [{\citenamefont {Regal}\ \emph {et~al.}(2004)\citenamefont {Regal},
  \citenamefont {Greiner},\ and\ \citenamefont {Jin}}]{Regal:2004zza}%
  \BibitemOpen
  \bibfield  {author} {\bibinfo {author} {\bibfnamefont {C.~A.}\ \bibnamefont
  {Regal}}, \bibinfo {author} {\bibfnamefont {M.}~\bibnamefont {Greiner}}, \
  and\ \bibinfo {author} {\bibfnamefont {D.~S.}\ \bibnamefont {Jin}},\ }\href
  {\doibase 10.1103/PhysRevLett.92.040403} {\bibfield  {journal} {\bibinfo
  {journal} {Phys. Rev. Lett.}\ }\textbf {\bibinfo {volume} {92}},\ \bibinfo
  {pages} {040403} (\bibinfo {year} {2004})}\BibitemShut {NoStop}%
\bibitem [{\citenamefont {Zwierlein}\ \emph {et~al.}(2004)\citenamefont
  {Zwierlein}, \citenamefont {Stan}, \citenamefont {Schunck}, \citenamefont
  {Raupach}, \citenamefont {Kerman},\ and\ \citenamefont
  {Ketterle}}]{Zwierlein:2004zz}%
  \BibitemOpen
  \bibfield  {author} {\bibinfo {author} {\bibfnamefont {M.~W.}\ \bibnamefont
  {Zwierlein}}, \bibinfo {author} {\bibfnamefont {C.~A.}\ \bibnamefont {Stan}},
  \bibinfo {author} {\bibfnamefont {C.~H.}\ \bibnamefont {Schunck}}, \bibinfo
  {author} {\bibfnamefont {S.~M.~F.}\ \bibnamefont {Raupach}}, \bibinfo
  {author} {\bibfnamefont {A.~J.}\ \bibnamefont {Kerman}}, \ and\ \bibinfo
  {author} {\bibfnamefont {W.}~\bibnamefont {Ketterle}},\ }\href {\doibase
  10.1103/PhysRevLett.92.120403} {\bibfield  {journal} {\bibinfo  {journal}
  {Phys. Rev. Lett.}\ }\textbf {\bibinfo {volume} {92}},\ \bibinfo {pages}
  {120403} (\bibinfo {year} {2004})}\BibitemShut {NoStop}%
\bibitem [{\citenamefont {Bedaque}\ \emph
  {et~al.}(1999{\natexlab{a}})\citenamefont {Bedaque}, \citenamefont {Hammer},\
  and\ \citenamefont {van Kolck}}]{Bedaque:1998kg}%
  \BibitemOpen
  \bibfield  {author} {\bibinfo {author} {\bibfnamefont {P.~F.}\ \bibnamefont
  {Bedaque}}, \bibinfo {author} {\bibfnamefont {H.~W.}\ \bibnamefont {Hammer}},
  \ and\ \bibinfo {author} {\bibfnamefont {U.}~\bibnamefont {van Kolck}},\
  }\href {\doibase 10.1103/PhysRevLett.82.463} {\bibfield  {journal} {\bibinfo
  {journal} {Phys. Rev. Lett.}\ }\textbf {\bibinfo {volume} {82}},\ \bibinfo
  {pages} {463} (\bibinfo {year} {1999}{\natexlab{a}})},\ \Eprint
  {http://arxiv.org/abs/nucl-th/9809025} {arXiv:nucl-th/9809025} \BibitemShut
  {NoStop}%
\bibitem [{\citenamefont {Bedaque}\ \emph
  {et~al.}(1999{\natexlab{b}})\citenamefont {Bedaque}, \citenamefont {Hammer},\
  and\ \citenamefont {van Kolck}}]{Bedaque:1998km}%
  \BibitemOpen
  \bibfield  {author} {\bibinfo {author} {\bibfnamefont {P.~F.}\ \bibnamefont
  {Bedaque}}, \bibinfo {author} {\bibfnamefont {H.~W.}\ \bibnamefont {Hammer}},
  \ and\ \bibinfo {author} {\bibfnamefont {U.}~\bibnamefont {van Kolck}},\
  }\href {\doibase 10.1016/S0375-9474(98)00650-2} {\bibfield  {journal}
  {\bibinfo  {journal} {Nucl. Phys.}\ }\textbf {\bibinfo {volume} {A646}},\
  \bibinfo {pages} {444} (\bibinfo {year} {1999}{\natexlab{b}})},\ \Eprint
  {http://arxiv.org/abs/nucl-th/9811046} {arXiv:nucl-th/9811046} \BibitemShut
  {NoStop}%
\bibitem [{\citenamefont {Geracie}(2017)}]{Geracie:2016dti}%
  \BibitemOpen
  \bibfield  {author} {\bibinfo {author} {\bibfnamefont {M.}~\bibnamefont
  {Geracie}},\ }\href {\doibase 10.1103/PhysRevB.95.134510} {\bibfield
  {journal} {\bibinfo  {journal} {Phys. Rev. B}\ }\textbf {\bibinfo {volume}
  {95}},\ \bibinfo {pages} {134510} (\bibinfo {year} {2017})},\ \Eprint
  {http://arxiv.org/abs/1612.01547} {arXiv:1612.01547} \BibitemShut {NoStop}%
\bibitem [{\citenamefont {Geracie}(2016)}]{Geracie:unpublished}%
  \BibitemOpen
  \bibfield  {author} {\bibinfo {author} {\bibfnamefont {M.}~\bibnamefont
  {Geracie}},\ }\href@noop {} {\bibfield  {journal} {\bibinfo  {journal}
  {unpublished}\ } (\bibinfo {year} {2016})}\BibitemShut {NoStop}%
\bibitem [{\citenamefont {Nishida}\ and\ \citenamefont
  {Son}(2007)}]{Nishida:2007pj}%
  \BibitemOpen
  \bibfield  {author} {\bibinfo {author} {\bibfnamefont {Y.}~\bibnamefont
  {Nishida}}\ and\ \bibinfo {author} {\bibfnamefont {D.~T.}\ \bibnamefont
  {Son}},\ }\href {\doibase 10.1103/PhysRevD.76.086004} {\bibfield  {journal}
  {\bibinfo  {journal} {Phys. Rev. D}\ }\textbf {\bibinfo {volume} {76}},\
  \bibinfo {pages} {086004} (\bibinfo {year} {2007})},\ \Eprint
  {http://arxiv.org/abs/0706.3746} {arXiv:0706.3746} \BibitemShut {NoStop}%
\bibitem [{\citenamefont {Nishida}\ and\ \citenamefont
  {Son}(2010)}]{Nishida:2009pg}%
  \BibitemOpen
  \bibfield  {author} {\bibinfo {author} {\bibfnamefont {Y.}~\bibnamefont
  {Nishida}}\ and\ \bibinfo {author} {\bibfnamefont {D.~T.}\ \bibnamefont
  {Son}},\ }\href {\doibase 10.1103/PhysRevA.82.043606} {\bibfield  {journal}
  {\bibinfo  {journal} {Phys. Rev. A}\ }\textbf {\bibinfo {volume} {82}},\
  \bibinfo {pages} {043606} (\bibinfo {year} {2010})},\ \Eprint
  {http://arxiv.org/abs/0908.2159} {arXiv:0908.2159} \BibitemShut {NoStop}%
\bibitem [{\citenamefont {Coutinho}\ \emph {et~al.}(1997)\citenamefont
  {Coutinho}, \citenamefont {Nogami},\ and\ \citenamefont
  {Fernando~Perez}}]{Coutinho:1997}%
  \BibitemOpen
  \bibfield  {author} {\bibinfo {author} {\bibfnamefont {F.}~\bibnamefont
  {Coutinho}}, \bibinfo {author} {\bibfnamefont {Y.}~\bibnamefont {Nogami}}, \
  and\ \bibinfo {author} {\bibfnamefont {J.}~\bibnamefont {Fernando~Perez}},\
  }\href {\doibase 10.1088/0305-4470/30/11/021} {\bibfield  {journal} {\bibinfo
   {journal} {J. Phys. A}\ }\textbf {\bibinfo {volume} {30}},\ \bibinfo {pages}
  {3937} (\bibinfo {year} {1997})}\BibitemShut {NoStop}%
\bibitem [{\citenamefont {Albeverio}\ \emph {et~al.}(2004)\citenamefont
  {Albeverio}, \citenamefont {Gesztesy}, \citenamefont {H\o{}egh-Krohn},\ and\
  \citenamefont {Holden}}]{Albeverio:2004}%
  \BibitemOpen
  \bibfield  {author} {\bibinfo {author} {\bibfnamefont {S.}~\bibnamefont
  {Albeverio}}, \bibinfo {author} {\bibfnamefont {F.}~\bibnamefont {Gesztesy}},
  \bibinfo {author} {\bibfnamefont {R.}~\bibnamefont {H\o{}egh-Krohn}}, \ and\
  \bibinfo {author} {\bibfnamefont {H.}~\bibnamefont {Holden}},\ }\href@noop {}
  {\emph {\bibinfo {title} {Solvable Models in Quantum Mechanics}}},\ \bibinfo
  {edition} {2nd}\ ed.\ (\bibinfo  {publisher} {American Mathematical
  Society},\ \bibinfo {year} {2004})\BibitemShut {NoStop}%
\end{thebibliography}%
\end{document}